\documentclass[aps,pra,twocolumn,amsmath,amssymb,showpacs]{revtex4}

\usepackage{graphicx}
\usepackage{dcolumn}

\begin{document}
\newcommand{\beq}{\begin{equation}}
\newcommand{\enq}{\end{equation}}
\newcommand{\lmax}{l_{\rm max}}
\newcommand{\lini}{l_{\rm i}}
\newcommand{\gat}{\gamma_{\rm at}}
\newcommand{\dele}{\Delta E_{\rm sc}}

\title{Radiative collisional heating at the Doppler limit for laser-cooled magnesium atoms}

\author{J. Piilo$^{1,2,3}$, E. Lundh$^{4,5}$, and K.-A. Suominen$^{1,4}$}

\affiliation{$^1$Department of Physics, University of Turku, FIN-20014 Turun yliopisto, Finland\\
$^2$Institute of Solid State Physics, Bulgarian Academy
of Sciences, Tsarigradsko chauss\'{e}e 72, 1784 Sofia, Bulgaria\\
$^3$School of Pure and Applied Physics, University of KwaZulu-Natal, Durban 4041, South Africa\\
$^4$Helsinki Institute of Physics, PL 64, FIN-00014 Helsingin yliopisto, Finland\\
$^5$Department of Physics, KTH, SE-10691 Stockholm, Sweden}

\date{\today}

\begin{abstract}
We report Monte Carlo wave function simulation results on cold collisions between magnesium atoms in a strong red-detuned laser field. This is the normal situation e.g. in magneto-optical traps (MOT). The Doppler limit heating rate due to radiative collisions is calculated for $^{24}$Mg atoms in a MOT based on the $^1$S$_0$-$^1$P$_1$ atomic laser cooling transition. We find that radiative heating does not seem to affect the Doppler limit in this case. We also describe a channelling mechanism due to the missing $Q$ branch in the excitation scheme, which could lead to a suppression of inelastic collisions, and find that this mechanism is not present in our simulation results due to the multistate character of the excitation process.

\end{abstract}


\maketitle

\section{Introduction} 

Laser cooling and trapping methods are an important ingredient in the recent achievements in the low-temperature physics of gaseous atoms and molecules~\cite{Metcalf03}. In alkali atoms one can reach very low temperatures with Sisyphus and polarization gradient techniques, which surpass the Doppler cooling method in efficiency, and appear in magneto-optical traps without additional efforts~\cite{Lett89}. The drawback is that it becomes impossible to test the basic two-state Doppler cooling theory with these systems. 

The appearance of sub-Doppler cooling is based on the hyperfine structure of the alkali atoms. The same hyperfine structure is reflected in the complicated molecular state structure of the quasimolecule formed by two colliding atoms. In dilute atomic gases binary collisions dominate the atomic interactions, and the quasimolecule states couple with the cooling and trapping light field. This leads to inelastic light-assisted collisions that cause either loss or heating of the trapped atoms~\cite{Suominen96,Suominen98,Weiner99}. Thus in alkali atoms the hyperfine structure leads to difficulties in testing the Doppler theory, as well as in modelling the collisional processes.

The major isotopes of alkaline earth atoms, on the other hand, have no nuclear spin and thus the hyperfine structure is missing. This makes the Doppler limit for temperatures the true laser cooling limit, and allows for testing the basic Doppler theory. Experimental studies have been published recently for $^{88}$Sr~\cite{Xu02}, and have been obtained for $^{24}$Mg as well~\cite{Thomsen03}. They indicate an intensity-dependent heating rate which prevents reaching the theoretical Doppler limit.

Inelastic laser-assisted collisions are expected to depend strongly on intensity even when the atomic transition is saturated ($I_s=$ 0.444 W/cm$^2$ for the $^1$S$_0$-$^1$P$_1$ laser cooling transition of $^{24}$Mg)~\cite{Suominen98}. In this region the problem is that most theoretical approaches are applicable only to the weak field situation, and can not handle the energy exchange when the energy change spectrum is continuous (as is the case for radiative collisional heating). An important exception is the Monte Carlo wave function method (MCWF)~\cite{Holland94,Piilo01}, which we apply here to the case of $^{24}$Mg, to obtain the average energy increase per collision near the Doppler limit (at 1 mK for $^{24}$Mg). Although simple in theory, the collisions between the alkaline earth atoms at strong fields are complicated since the partial waves and rotational states of the ground and excited states (respectively) of the quasimolecule form a network of two infinite sets of coupled states~\cite{Machholm01}.   

In this article we report the results of Monte Carlo simulations of collisional radiative heating, based on a truncated set of partial waves and rotational states. We derive a heating rate due to these collisions, and compare it to the photon scattering heating rate that sets the Doppler limit in the standard collisionless theory. We also outline a channelling mechanism which is possible due to the specific structure and selection rules of the alkaline earth dimer quasimolecule, and which could lead to the suppression of inelastic processes, and test it with the simulations. Here we concentrate on the results provided by the simulations, and report the multitude of technical simulation details elsewhere.

\section{Magnesium quasimolecule} 

The lasers for cooling and trapping in magneto-optical traps (MOT) are normally detuned a few atomic linewidths ($\gat$) below the atomic transition, which means that the attractive quasimolecule states have a finite internuclear distance, the Condon point $R_C$, where they become resonant with the laser field. For $^{24}$Mg and other alkaline earth atoms we have two such states, a strongly coupled $^1\Sigma_u^+$ state, and a weakly coupled $^1\Pi_g$ state~\cite{Stevens77,Czuchaj01}. 

In the radiative heating mechanism, the colliding atoms on a partial wave $l$ are excited at $R_C$ to an attractive quasimolecule state with rotational quantum number $J$. The kinetic energy of the relative atomic motion increases as the atoms approach each other further, until they return to the ground state via spontaneous emission. The energy exchange due to such a process forms a continuous distribution below the trap depth energy (above which the radiative heating becomes radiative escape). 

\begin{figure}[th]
\centering
{\includegraphics[width=\columnwidth]{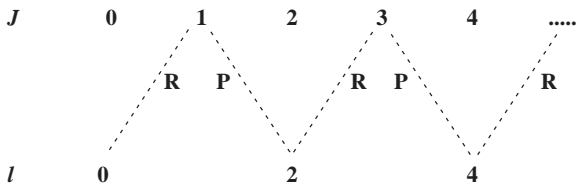}}
\caption[f1]{The coupling scheme for the $^1\Sigma_g^+$ ground state partial wave $l$ and the $^1\Sigma_u^+$ excited state rotational state $J$.\label{fig:Branching}}
\end{figure}

Since the $^1\Pi_g$ state is only weakly coupled to the light field, and provides mainly a channel for trap loss, we can concentrate in our modelling of heating on the $^1\Sigma_u^+$ state only (see Ref.~\cite{Machholm01} for a more detailed discussion). The ground state partial waves and the excited rotational states are coupled by the selection rules $l\rightarrow J=l\pm 1$, i.e., the $Q$ branch for molecular excitation is missing. In the weak field limit one can consider each partial wave as paired independently to a rotational state, forming a set of independent two-state systems. At strong fields multiple couplings are possible, and one can move back and forth on the ``angular momentum ladder'' shown in Fig.~\ref{fig:Branching}. Note that for the bosonic $^{24}$Mg atoms the exchange symmetry allows only even values for $l$.

\begin{figure}[th]
\centering
\scalebox{0.9}
{\includegraphics[width=\columnwidth]{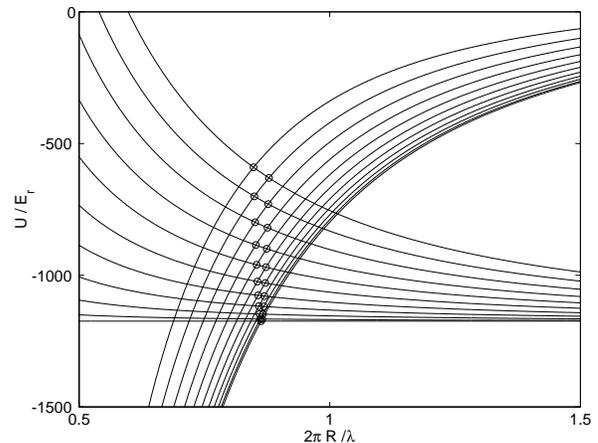}}
\caption[f2]{\label{fig:Potentials}
Ground and excited state potentials for partial waves/rotational states
up to $l=20$ and $J=21$. Open circles indicate the point of couplings forming a net of Condon points. Here $\delta=-3\gat$. The ground states have been shifted up by the energy of one photon.}
\end{figure}

{\it Channelling.} One should note, as Fig.~\ref{fig:Potentials} illustrates, that each pair ($l,J$) is coupled resonantly at a {\it different} value of $R_C$. Thus at modest field strengths one could consider a situation where the separation of the Condon points is valid while the excitation probability is close to unity. Due to the missing $Q$ branch one can see that except for the $s$-wave, we always have a sequence $l\rightarrow J=l-1\rightarrow l'=J-1=l-2$ so that an incoming ground state partial wave $l$ is channelled via two crossings into another ground state partial wave $l''=l-2$, then reflected by the centrifugal barrier, returning the same route back to the original state $l$. This process would reduce the collisions practically elastic (the $s$-wave still remains "unchannelled", though).   

\section{Collisional heating rate} 

The energy increase from a single collision connects to a heating rate and the appropriate rate coefficient in the following way. The collisional heating rate $\kappa_H (n)$ describes the kinetic energy change per unit volume and unit time. For identical particles it will be equal to $\frac{1}{2}K_Hn^2$, where $K_H$ is the rate coefficient for collisional heating, and $n$ is the atomic density. The factor of $1/2$ removes the doubling in collision counting for identical particles. The total collisional heating rate is obtained by integrating $\kappa_H$ over the trap volume.

For trap loss rate coefficient $K_{\rm loss}$ one normally calculates the collision frequency times the probability for loss, i.e., $v\sigma_{\rm loss}$, where $v$ is the relative velocity, and $\sigma_{\rm loss}$ is the cross-section for collisions, including the probability for a loss event to occur. This simple classical picture is connected to thermodynamics by assuming a thermal equilibrium distribution of velocities, $f(v)dv$, over which we take an average, and obtain $K_{\rm loss}=\langle v\sigma_{\rm loss}\rangle$. The dynamics of the two-body collision on the microscopic level then enters in calculating $\sigma_{\rm loss}$.

To calculate the heating rate coefficient $K_H$ is slightly more complicated, because in addition to the probability of an inelastic collision event to happen, we also need to estimate the amount of kinetic energy increase associated with it. Technically, we would have a continuous distribution of final energies corresponding to each initial value of $v$, due to the randomness of the spontaneous emission events. For practical reasons we consider an averaged rate, i.e., we calculate the average change in kinetic energy per unit time. Since we can perform the averaging over final energy states before the averaging over initial states, we define an inelastic (heating) cross section $\sigma_H(v)$ (units energy$\times$length$^2$) that gives the difference of the average final relative kinetic energy and initial relative kinetic energy.

In the partial wave approximation we can write the quantum mechanical cross-section for identical atoms in a three-dimensional trap as~\cite{Julienne89}
\begin{equation}
   \sigma(v) =\frac{\pi}{k^2}\sum_{l=0 {\rm (even)}}^{\infty} (2l+1)P_l(v),
\end{equation}
where $k$ is the wave number related to $v$ and $P_l(v)$ is the event probability. Thus, as a generalization, we write for heating
\begin{equation}
   \sigma_H(v) =\frac{\pi}{k^2}\sum_{l=0 {\rm (even)}}^{\infty} 
   (2l+1)\dele(v,l),
\end{equation}
where $\dele(v,l)$ is the average single-collision energy increase related to the initial partial wave $l$. 

In principle one should take the thermal average over an isotropic Maxwell-Boltzmann distribution of relative velocities $v$. Due to the complexity of the Monte Carlo simulations we are, in practice, limited to calculating $\dele(v,l)$ for a rather narrow range of initial relative velocities $v$. As in Ref.~\cite{Machholm01}, we define a single-energy rate coefficient as ($E=\frac{1}{2}\mu v^2$)
\begin{equation} 
   K_H(E) = \frac{E}{hQ_E}
   \sum_{l=0 {\rm (even)}}^{\infty} (2l+1) \dele(v,l),\label{KHE}
\end{equation}
with $Q_E=(2\pi\mu E/h^2)^{3/2}$, where $\mu$ is the reduced two-particle mass. The simulations have shown this to be a reasonable approximation.

\section{Simulation results} 

We have performed a series of computationally tedious simulations at the detuning $\delta = -3\gat$ for various laser intensities, and the main results are given in Table~\ref{table1}. We have used the semiclassical trajectory argument (turning point should be at $R<R_C$ for any involved partial wave) to truncate the sum in Eq.~(\ref{KHE}). Clearly a strong increase in the energy change with intensity continues above the saturation intensity $I_s$. The semiclassical Landau-Zener theory also indicates that in this intensity region excitation probabilities should be equal to unity. Thus the continued increase can be attributed to the re-excitation of the decayed population, as suggested in Ref.~\cite{Suominen98}; the delayed decay model predicts (roughly) $\dele\propto\sqrt{I/I_s}$.

\begin{table}[th]
\begin{tabular}{rrcc}
\hline
\hline
$I$ (W/cm$^2$) & $I/I_s$ & $\delta/\gat$ & $\dele/E_r$\\
\hline
0.036 &  0.08 & -3.0  &  16 $\pm$  3 \\
0.88  &  2.0  & -3.0  & 145 $\pm$ 16 \\
2.2   &  5.0  & -3.0  & 249 $\pm$ 33 \\
3.6   &  8.0  & -3.0  & 307 $\pm$ 31 \\
5.3   &  12.0 & -3.0  & 394 $\pm$ 42 \\
8.0   &  18.0 & -3.0  & 427 $\pm$ 40 \\
14.2  &  32.0 & -3.0  & 547 $\pm$ 81 \\
\hline
\hline
\end{tabular}
\caption[t1]{\label{table1}
Results from the multistate single-collision simulations. The total number of states has been 12, and the initial state has been $l=8$. The energy increase $\dele$ is calculated as a time average of the kinetic energy in the region where it is flat, subtracted by the initial energy. The error refers to the statistical error of the average kinetic energy at the end of the Monte Carlo simulation.}
\end{table}

Additional simulations showed that pre- and post-collision steady state formation is fast and strong for $I\gtrsim 5I_s$, and makes the energy exchange independent of the initial partial wave, i.e., $\dele(v,l)\rightarrow\dele(v)$ in Eq.~(\ref{KHE}). In other words, in a system where the energy states are strongly coupled and remain so even asymptotically, the definition of an initial state is obscured, and we can actually make meaningful studies because the spontaneous emission always leads to the same steady state, independent of initial $l$, well before $R_C$ is reached.  For $I\lesssim 5I_s$ there is some initial state dependence, but on a scale that allows one to use the values given in Table~\ref{table1} as a reasonable estimate of magnitude.

\begin{figure}[th]
\centering
\scalebox{0.9}
{\includegraphics[width=\columnwidth]{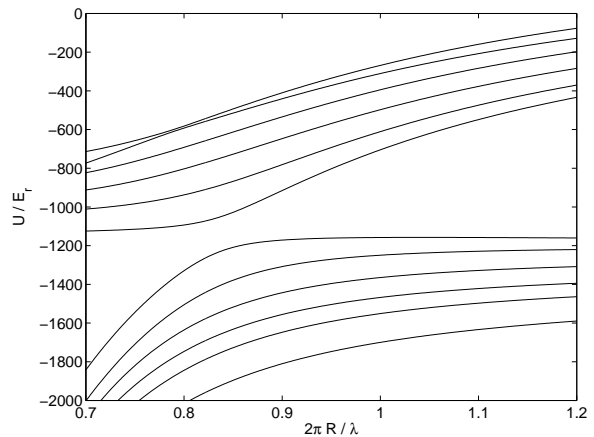}}
\caption[f3]{\label{fig3}
The field-dressed potentials for the magnesium quasimolecule
up to $l=12$ and $J=13$. Here $\delta=-3\gat$ and $I/I_s=8.0$. }
\end{figure}

As for the channelling effect, there was no sign of it in the simulations. The explanation is rather clear. The Condon points are close to each other both in position and energy, and the true picture of the excitation process is more like adiabatic following of the field-dressed (adiabatic) quasimolecule states, shown in Fig.~\ref{fig3}. However, this may not be the case for other values of detuning, or weaker fields than those used in our simulations. For this reason we want to emphasise the general possibility for channelling, and also because it is a good test for the validity of the reduction into independent two-state models~\cite{Machholm01}. We have confined our computationally demanding simulations to large intensities because then there are suitably many quantum jumps for having reasonable statistics for small ensembles of 64 or 128 members.

Finally it should be noted that any truncation of the partial wave/rotational state manifolds must be done so that the total number of states is even. Otherwise one would introduce an artificial dark state as one of the field-dressed eigenstates, into which the strong and fast steady state formation would trap the system well before the actual collision begins.

\section{Comparison of rates} 

The heating rate prediction of the simulations must be compared to the photon scattering heating rate that enters the calculation of the Doppler limit. In this we follow Ref.~\cite{Lett89} and obtain the approximate rate
\begin{equation} 
  \left(\frac{dE}{dt}\right)_{\rm heat}
  =\frac{E_r\gat}{2} \dfrac{I/I_s}{1+I/I_s+(2\delta/\gat)^2},
\end{equation}
where $E_r=\hbar^2k_r^2/(2\mu)$ is the photon recoil energy, and $I_s=\pi \gat h c/(3\lambda^3)$ is the saturation intensity~\cite{Metcalf03}. Adding the numbers for $^{24}$Mg, we get the heating rate per particle ($I\simeq I_s$)
\begin{equation} 
   \left(\frac{dE}{dt}\right)_{\rm heat} \simeq \dfrac{E_r\gat}{80} 
   \simeq 60 \ \frac{\rm K}{\rm s}.\label{SPR}
\end{equation}
The above expression has been derived for a one-dimensional two-beam setup, and its extension to the full six-beam three-dimensional situation is an interesting and open question, and one of the strong motivations to study Doppler cooling with alkaline earth atoms.

The collisional heating rate coefficient becomes, using Eq.~(\ref{KHE}),
\begin{equation} 
  K_H(E) = 4\pi^{3/2}\lambdabar^3\dfrac{E_r^2}{h}\left(\dfrac{E}{E_r}\right)^{1/2}
  \left(\dfrac{3\gat}{2|\delta|}\right)^{2/3}\dfrac{\dele}{E_r}.
\end{equation}
Here $\lambdabar=k_r^{-1}=\lambda/(2\pi)$, with $\lambda=285.21$ nm for $^{24}$Mg. This result is obtained if we note that by truncating the partial wave series to the maximum classically allowed $l$, denoted by $l_{\rm max}$, we have $E=\hbar^2 l_{\rm max}(l_{\rm max}+1)/(2\mu R_C^2)$. The Condon point $R_C$ is determined by the $^1\Sigma_u^+$ state potential, $U(R)=-3\hbar\gat/[2(k_rR)^3]$, being equal to $\hbar\delta$ at $R_C$. Finally, $\sum_{l=0 {\rm (even)}}^{l_{\rm max}}(2l+1)=(l_{\rm max}+1)(l_{\rm max}+2)/2\simeq l_{\rm max}(l_{\rm max}+1)/2= \mu R_C^2 E/\hbar^2$. To compare with the single-particle rate (\ref{SPR}) we write (for $\delta = -3\gat$)  
\begin{equation} 
   \left(\frac{dE}{dt}\right)_{\rm coll} =\frac{1}{2}K_H(E)n\simeq k_B\times 1.3
   \times 10^{-14}\ \dfrac{\rm K}{\rm s}
   \times\dfrac{\dele}{E_r}\times n.
\end{equation}
For $n\simeq 10^{11}$ atoms/cm$^3$ and $I\simeq I_s$ this gives about 0.1 K/s. This result implies that radiative collisional heating can not explain the problems in reaching the Doppler limit for $^{24}$Mg, because the densities are usually in the range of $10^9$ to $10^{11}$ atoms/cm$^3$. We have earlier performed two-state studies for Cs and found that there the heating rate present in Sisyphus cooling should match the collisional heating rate at $n= 10^{12}$ atoms/cm$^3$~\cite{Holland94}. 

As for making a comparison to the $^{88}$Sr case, without performing simulations, we can make a rough estimate. In Eq.~(\ref{KHE}) we have written $K_H(E)$ in a scaled form, so we can put in the corresponding values for $^{88}$Sr as well ($E/k_B\simeq T_D =0.77$ mK, $\lambda= 460.73$ nm, $E_r=1\ \mu$K, and taking again $|\delta|=3\gat$). The remaining problem is to estimate $\dele$. In a toy model that assumes total excitation during a collision, and ignores re-excitation, we can write $\dele\simeq U'(R_C)\tau v$, where $U'(R_C)$ is the slope of the excited state (ignoring $J$) potential at $R_C$, and $\tau\simeq 1/\gat$ is the average survival time on the excited molecular state, and $v$ is the initial collision velocity. This leads to an estimate of $\dele/E_r=23(E/E_r)^{1/2}$, which at least for $^{24}$Mg gives 230, i.e., the correct magnitude near $I=I_s$. With all this, we get for $^{88}$Sr at the Doppler temperature the result $K_H(E)\simeq 2\times 10^{-12}$ Kcm$^3$/s, to be compared with $E_r\gat/80\simeq 2.5$ K/s. These results meet only at densities of $n\simeq 10^{12}$ atoms/cm$^3$, which are a few magnitudes higher than those used in the experiment of Ref.~\cite{Xu02}. This toy model also predicts a detuning dependence of $K_H(E)$ of $\propto|\delta|^{2/3}$, i.e., heating decreases with detuning, which means that while the number of involved partial waves increases with diminishing $|\delta|$, the effect is more than countered by the reduction in the steepness of the excited state potential. 

\section{Conclusions} 

This study is the first attempt to describe quantitatively the radiative collisional heating by taking into account the inherent multistate structure due to the partial waves and rotational states. The truncation of the partial wave manifold has required careful testing of various possibilities. Although it appears that collisional radiative heating does not play a role in the thermodynamics of the trap at realistic densities for $^{24}$Mg (and other alkaline earth atoms), our work does not rule it out completely. Although assumed to be irrelevant here, the weakly coupled $^1\Pi_g$ state may play a role, as well as the two quasimolecule states corresponding to the repulsive potentials. Experimentally the role of collisional processes can be distinguished from other processes by looking at the scaling with density; this has already been used to separate the cold collision contributions from the background collisions in trap loss. While we aim to improve our modelling and to use our heating results in estimating radiative trap loss, we also expect that more experimental data will become accessible in the near future. It should be pointed out that other effects may also lead to heating. The methods to study e.g. radiative trapping (reabsorption of scattered photons) are, however, very different from the ones employed here. 

\section{Acknowledgments} 

The authors acknowledge the EU network CAUAC (contract HPRN-CT-2000-00165) and the Academy of Finland (projects 50314 and 206108) for financial support. JP has also been supported by the EU network COCOMO (contract HPRN-CT-1999-00129) and the Academy of Finland (project 204777). Discussions with E. Rasel, J. Thomsen and J. Ye are gratefully acknowledged, as well as useful comments by I.~Mazets and J.~Vigu\'{e}. The Finnish IT Center for Science (CSC) is gratefully acknowledged for providing the supercomputer environment used in the simulations.

\end{document}